\documentclass[aps,preprint,floatfix,superscriptaddress]{revtex4-2}
\usepackage{graphicx}
\usepackage[colorlinks=true,linkcolor=blue,citecolor=blue,urlcolor=black]{hyperref}
\usepackage{xcolor}
\usepackage{etoolbox}
\usepackage{amsmath}
\usepackage[capitalize]{cleveref}
\def\beq{\begin{equation}}
\def\eeq{\end{equation}}
\def\bea{\begin{eqnarray}}
\def\eea{\end{eqnarray}}
\newrobustcmd*{\mycircle}[1]{\tikz{\filldraw[draw=#1,fill=#1] (0,0) circle [radius=0.1cm];}}
\definecolor{col1}{rgb}{0.4 0.5 0.8}
\definecolor{col2}{rgb}{1 0.7 0}

\begin{document}
\makeatletter
\title{An alternate approach to simulate the dynamics of perturbed liquid drops}

\author{Tanu Singla}
\email{tanu.singla@tec.mx}
\affiliation{Tecnologico de Monterrey, Calle del Puente 222, Colonia Ejidos de Huipulco, Tlalpan, Ciudad de M\'exico, M\'{e}xico}

\author{Tanushree Roy}
\affiliation{Department of Physics,
Indian Institute of Technology Bombay,
Powai, Mumbai-400 076, India}

\author{P. Parmananda}
\affiliation{Department of Physics,
Indian Institute of Technology Bombay,
Powai, Mumbai-400 076, India}

\author{M. Rivera}
\affiliation{Centro de Investigaci\'on en Ciencias-(IICBA), UAEM, Avenida Universidad 1001, Colonia Chamilpa, Cuernavaca, Morelos, M\'{e}xico}

\date{\today}

\begin{abstract}

Liquid drops when subjected to external periodic perturbations can execute polygonal oscillations. In this work, a simple model is presented that demonstrates these oscillations and their characteristic properties. The model consists of a spring-mass network such that the masses are analogous to liquid molecules and the springs are to intermolecular forces. Neo-Hookean springs are considered to represent these intermolecular forces. The restoring force of a neo-Hookean spring depends nonlinearly on its length such that the force of a compressed spring is much higher than the force of a spring elongated by the same amount. This is equivalent to the incompressibility of liquids, making these springs suitable to simulate the polygonal oscillations. It is shown that this spring-mass network can imitate most of the characteristic features of experimentally reported polygonal oscillations. Additionally, it is shown that the network can execute certain dynamics which so far have not been observed in a perturbed liquid drop. The features of dynamics which are observed in the perturbed network are: polygonal oscillations, rotation of network, numerical relations (rational and irrational) between the frequencies of polygonal oscillations and the forcing signal, and the dependency of the shape of the polygons on the parameters of perturbation.
\end{abstract}

\maketitle

{\bf Perturbation of a liquid drop implies forcing it with an external signal. When this signal is periodic in nature, the perturbed drop can oscillate in the form of polygons. In the past, different types of perturbation signals (mechanical, electrical, chemical, thermal, etc.) have been used and all of these result in similar polygonal oscillations in the liquid drops. In this work, we are presenting a numerical model which shows oscillations identical to those observed in experimentally perturbed drops. For this purpose, a network of springs and masses is developed and its governing equations are solved numerically. It is shown that this network is able to capture most of the characteristic properties of the dynamical behavior of the perturbed liquid drops.}

\section{Introduction}
The physics behind the dynamics of liquid drops has been a subject of interest among the scientific community for a long time. Leidenfrost effect, where a liquid drop when kept on a surface at a temperature higher than the boiling point of liquid hovers over its vapor layer, is known to the scientific community since the 1750s \cite{Leidenfrost}. Using the Marangoni effect, the interplay between surface tension and evaporation of a liquid drop can explain numerous phenomena; Marangoni introduced this effect in the 1890s in his pioneering work on ``Marangoni flows'' \cite{Marangoni}. With the dawn of technology, scientists have designed novel techniques to study liquid drops and numerous phenomena are being reported. This includes revealing rotational nature of a Leidenfrost drop \cite{Bouillant}, the splash dynamics upon the impact of a falling drop on a surface \cite{Santiago}, walking and orbiting droplets on a liquid surface \cite{Couder}, superpropulsion of droplets and soft elastic solids \cite{Raufaste} by using high speed video recording cameras and collapse of air films on drop impact \cite{Jolet}, and Marangoni bursts \cite{Keiser} with the use of interferometric techniques. Other phenomena that have been reported in liquid drops are vapor-mediated sensing and motility in two-component droplets \cite{Cira}, internal dynamics during the coalescence of a sessile droplet \cite{Sykes}, impact of a falling drop on solid surfaces \cite{Mostafa}, synchronization \cite{Verma,Dinesh1,Tanu,Pawan,Pawan1}, cessation \cite{Richa} and control \cite{Pawan2} of oscillations.

Liquid drops when subjected to external periodic perturbations, execute polygonal oscillations. The pioneering work on such oscillations in liquid drops was done by Holter et al. \cite{Holter} where they reported these oscillations in a Leidenfrost drop. This work has recently been extended by Ma et al. \cite{Ma} in which polygonal oscillations in Leidenfrost drops of different liquids are reported. These oscillations have also been observed in a vertically vibrating water drop \cite{Yoshiyasu}, in a magnetic fluid drop kept in a static magnetic field \cite{Jamin}, in acoustically levitated drops \cite{Yarin,Shen}, and in drops levitated by airflow \cite {Bouwhuis}. Our group has also studied these oscillations using electrical \cite{Dinesh,Elizeth}, mechanical \cite{Singla}, and thermal \cite{Singla1} perturbations of liquid drops. The major observations reported in these works are that the number of lobes in the polygons of the liquid drops is a function of the perturbation frequency, the shape of the polygon changes periodically in every cycle of the perturbation, and the drop at certain parameters can execute rotational motion. The fact that the shape of the polygons in the drop changes in every cycle leads to the existence of specific numerical relations between the frequency of oscillations of the drop ($f_d$) and the frequency of perturbation ($f_p$). In most of the previous studies, it has been shown that $f_d$ and $f_p$ are related by a factor of half ($f_d=f_p/2$). Recently, it has been been reported that at some perturbation frequencies, $f_d$, and $f_p$ can also be related by a factor of one-quarter ($f_d=f_p/4$) \cite{Singla}.

Theoretical and numerical investigations of this behavior have also been attempted by different groups. The authors of \cite{Shen,Okada} used predefined mathematical functions to approximate the solutions of the oscillations. Models involving partial differential equations are used in \cite{Bouwhuis,Pototsky} to study shapes of drops undergoing periodic vibrations. Recently, linear stability analysis has also been performed on a system involving a radially perturbed viscous droplet immersed in another inviscid fluid \cite{Yikai}. While these studies successfully demonstrated several important characteristics of the oscillations of the perturbed drop, they failed to capture the polygonal nature of the shapes that is observed experimentally.

The dynamics of liquid drops in different situations have also been modelled by considering them as a system of springs and masses. For e.g., deformation and secondary breakup of a drop kept in a gaseous flow \cite{Marek} and bouncing of liquid drops on different surfaces \cite{Terwagne,Molacek,Blanchette,Blanchette1} have been modelled using linear and nonlinear springs. In this work, a simple model that imitates polygonal oscillations in a perturbed liquid drop is presented. The model is composed of a network of multiple masses interconnected with neo-Hookean springs. Neo-Hookean springs are characterized by nonlinear restoring force such that the force of a compressed spring is higher than that of the spring elongated by the same length. This makes neo-Hookean springs suitable to mimic the incompressibility of liquid drops. This is also evident from a recent study where the bounces of hydrogel spheres ($\sim$99\% water) were modeled by using a linear chain of these springs and masses \cite{waitukaitis}. It is shown that the perturbed network can execute polygonal oscillations and that the shape of the polygons depends on the parameters of the perturbation. Different quantitative relations between the frequency of the oscillations in the network ($f_n$) and that of the perturbation ($f_p$) are also established in our calculations.

\section{Numerical model}\label{sec2}

\begin{figure}[ht!]
\includegraphics[trim=0cm 2cm 0cm 1cm,clip=true,width=10cm,height=7cm]{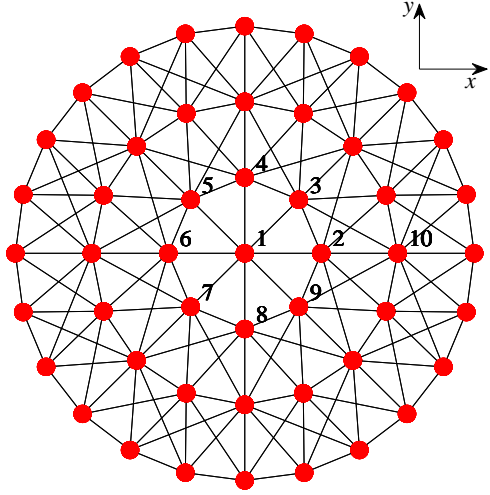}
\caption{Schematic representation of the spring-mass network used to study polygonal oscillations. Red circles represent masses and black lines represent neo-Hookean springs.}
\label{1}
\end{figure}

A schematic diagram of the spring-mass network is shown in \cref{1}; the red circles represent masses and the lines represent neo-Hookean springs. The network is constrained to move on a horizontal plane; gravitational force on the masses, therefore, is not incorporated in the calculations. The restoring force of a neo-Hookean spring is given by the following relation:

\begin{alignat}{2}
F_r&=-k\Big(\lambda-\frac{1}{\lambda^2}\Big),\label{eq1}\\
\lambda&=\frac{\Delta}{\delta}.
\end{alignat}
Here, $k$, and $\delta$ respectively are the spring constant, and the resting length of the springs. $\Delta$ represents the distance between two masses connected by a spring and can be calculated using the coordinates of the masses with the relation: $\Delta=[(x_1-x_2)^2+(y_1-y_2)^2]^{1/2}$. 

The effect of damping is also considered in our network. A cubic dependence of the damping force on the velocity is considered and is given by the following relation:

\begin{alignat}{2}
F_d&=-bv_r^3,\label{eq2}\\
v_r&=[(v_{x_1}-v_{x_2})^2+(v_{y_1}-v_{y_2})^2]^{1/2}.
\end{alignat}
Here, $b$ is the damping constant of the spring, and $v_r$ is the relative velocity of two masses connected by the spring. It needs to be pointed out that in the initial calculations, a linear dependence of $F_d$ on $v_r$ was implemented and it did not result in the polygonal oscillations. Therefore, the next choice was to incorporate nonlinearity in the damping term. As the direction of the damping force of a spring is always opposite to that of the velocity, and also to keep the nonlinear term simple, a cubic dependence was implemented. 

The governing differential equations for the perturbed spring-mass network can then be written using \cref{eq1}, \cref{eq2}, a perturbation term (shown later), and the information of coupling relations between various masses. To obtain these coupling relations, it is convenient to first assign numbers to the masses and then write an adjacency matrix ($A$)  providing information about the links between the masses. Masses in the network in \cref{1} are considered to be arranged in four layers; mass at the center being the first layer. The mass at the center is assigned number $1$, the counting then continues for masses of the second layer and moving outwards; this numbering scheme is illustrated in \cref{1}. Elements of the adjacency matrix are: $A_{ij}=1$ if $i^{th}$ and $j^{th}$ masses are connected and $0$ otherwise. Following the above mentioned protocol, the dynamics of $i^{th}$ mass in the perturbed network can be expressed with the following relation:

\begin{equation}\label{eq3}
m\frac{d^2{\bf x_i}}{dt^2}=\sum_{j=1}^N A_{ij}\Big[{\bf F_{r_{ij}}}+{\bf F_{d_{ij}}}\Big]+a\mathrm{sin}(2\pi f_pt){\bf c_{\theta_i}}.
\end{equation}
Here, $m$, and ${\bf x_i}=\{x_i,y_i\}$ respectively are mass and position coordinates of the $i^{th}$ mass, ${\bf F_{r_{ij}}}$ (\cref{eq1}), and ${\bf F_{d_{ij}}}$ (\cref{eq2}) respectively are the restoring and damping force of the spring connecting $i^{th}$ and $j^{th}$ masses, $a$ and $f_p$ are the amplitude and the frequency of external perturbation on the network, and ${\bf c_{\theta_i}}=\{\mathrm{cos}\theta_{i},\mathrm{sin}\theta_{i}\}$ is the unit vector in the direction of perturbation ($a\mathrm{sin}(2\pi f_pt)$); $\theta_i$ being the angle which the $i^{th}$ mass makes with the $x$-axis at the origin. If $n_{l_j}$ ($j=\{1~2~3~4\}$) represents number of masses in each layer of the spring-mass network (\cref{1}), then $n_{l_1}=1$, $n_{l_2}=8$, $n_{l_3}=2n_{l_2}$, and $n_{l_4}=3n_{l_2}$ is considered; $N=\sum_jn_{l_j}$ ($N$ is the total number of masses in the network). To avoid translational motion of the network, $m_1$ is fixed at the origin.

Apart from the perturbation term considered in \cref{eq3}, the dynamics of the network depend on three parameters ($k$, $b$, and $\delta$). To facilitate selection of appropriate values of these parameters, \cref{eq3} is nondimensionalized by considering ${\bf x_i}=x_0{\bf x_i'}$ and $t=t_0t^{'}$. With this, \cref{eq3} transforms to:

\begin{equation}
\label{eq4}
\frac{d^2{\bf x_i'}}{dt^{'2}}=\sum_{j=1}^N A_{ij}\Big[{\bf F^{'}_{r_{ij}}}+{\bf F^{'}_{d_{ij}}}\Big]+a'\mathrm{sin}(2\pi f_p^{'}t^{'}){\bf c_{\theta_i}}.
\end{equation}
For simplicity, the detailed procedure of nondimensionalization of \cref{eq3} is provided in the \cref{app1}. In \cref{eq4}, the restoring force (${\bf F^{'}_{r_{ij}}}$) is parameter independent and the damping term (${\bf F^{'}_{d_{ij}}}$) depends only on one parameter ($b^{'}$); making the nonperturbed network also dependent on this parameter $b^{'}$.

\section{Analysis and Terminology}\label{sec3}
{\bf Analysis:} For simulation results, \cref{eq4} is solved using the RK4 algorithm with a step size of 0.001, $b^{'}=1$ and different values of $a'$ and $f_p^{'}$. For each set of ($a'$, $f_p^{'}$) values, time series for the $x'$ and $y'$ coordinates of all the masses in the network is obtained. Among other techniques for the analysis, the principal ones that are employed on the time series are to plot the instantaneous positions of the masses (in $x'-y'$ space), and frequency spectra of the time series of one of the masses and that of the perturbation signal.

{\bf Terminology:} {\em Topology} of a network is the physical distribution of its masses in the $x'-y'$ space.  A network will be called to be executing {\em symmetric oscillations} if at any time instant, all of its masses are equally distributed among four quadrants of the $x'-y'$ space. On the contrary, topologies of the network with unequal distribution of masses will constitute {\em asymmetric oscillations}. Finally, two polygons will be called {\em inverted polygons} if in one polygonal shape, the radius of a mass is larger than that of another in the same layer of the network. Consequently, in the inverted polygon radius of first mass will become smaller than that of the other.

\section{Numerical Results}\label{sec4}

\begin{figure*}[ht!]
\includegraphics[width=16cm,height=8cm]{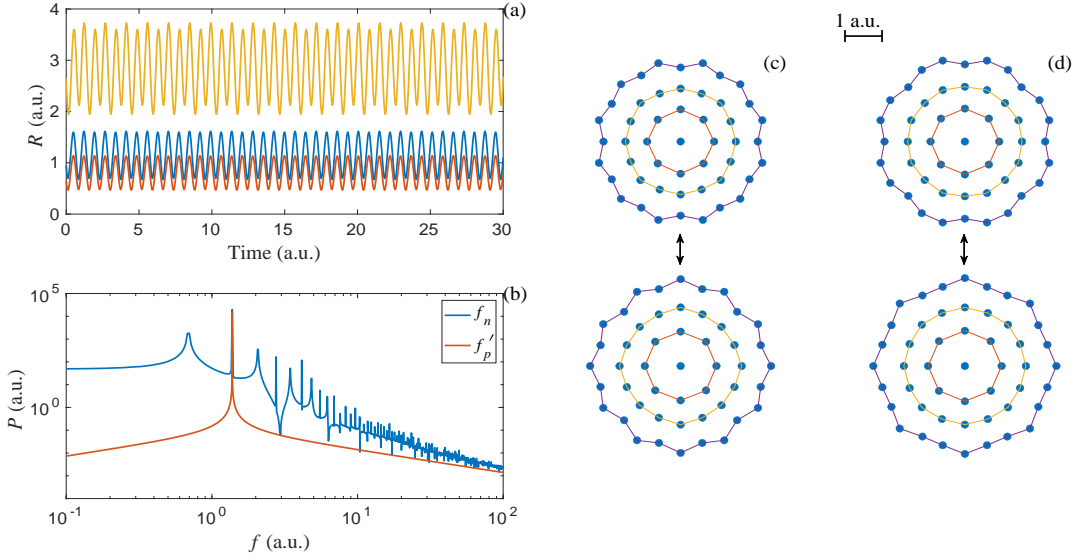}
\caption{(a) Time series of the radii of a single mass from outermost three layers (layers 2, 3 , and 4) of the network, (b) frequency spectra of a mass from the outermost layer (blue curve) and the perturbation signal (orange curve) for $a'=4.8$, $f_p^{'}=1.38$. Inverted polygonal shapes in the network for (c) $a'=4.8$, $f_p^{'}=1.38$ (multimedia view), and (d) $a'=3.6$, $f_p^{'}=1.38$ (multimedia view).}
\label{2}
\end{figure*}

\begin{figure}[ht!]
\includegraphics[width=16cm,height=8cm]{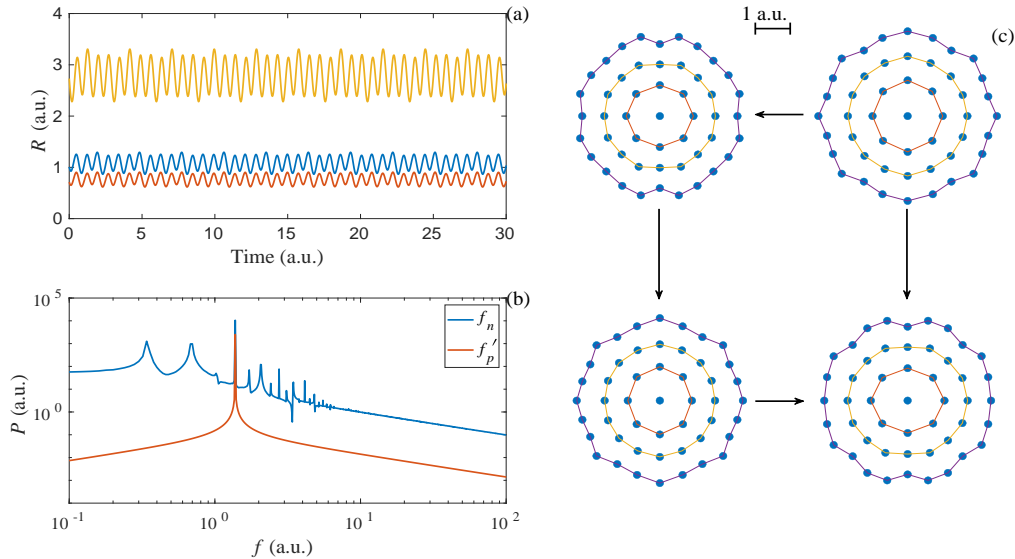}
\caption{(a) Time series of the radii of one mass from outermost three layers of the network, (b) frequency spectra of a mass from the outermost layer (blue curve) and the perturbation signal (orange curve), and (c) inverted polygonal shapes in the network for $a'=2.2$, $f_p^{'}=1.38$ (multimedia view).}
\label{3}
\end{figure}

\cref{2} shows the time series of the radii of masses, its frequency spectra, and the topology of the network for different perturbation parameters. \cref{2}(a) represents the time series of three masses belonging to the outermost three layers of the network ($m_1$ is fixed at the origin, so its time series is not shown). The frequency spectra of the oscillations of the network (blue curve) along with that of the perturbation (orange curve) are shown in \cref{2}(b). It can be observed that the frequency of the oscillations of the network is half of the perturbation frequency ($f_n=f_p^{'}/2$). This is in line with the experimental results reported in several works on the polygonal oscillations in perturbed liquid drops \cite{Holter,Ma,Yoshiyasu,Jamin,Shen,Bouwhuis}. Moreover, it is known that the topology of the polygons of a perturbed drop inverts during every cycle of the perturbation. This behavior is demonstrated in \cref{2}(c) (multimedia view) where two inverted polygons of the spring-mass network corresponding to time series in \cref{2}(a) are shown. Arrow marked between the polygons indicates that the network oscillates between these two shapes. Moreover, as defined in the previous section and on the basis of the topology of the polygons in this case, oscillations in the network are symmetric in nature. Inverted polygons satisfying the relation $f_n=f_p^{'}/2$ are provided for another set of perturbation parameters in \cref{2}(d) (multimedia view). Results of \cref{2}((c) and (d)) show that as the perturbation parameters change, shapes of the polygons observed in the network also change, which again is in accordance with the previously reported results. All of these observations from \cref{2} indicate that the oscillations in the spring-mass network are qualitatively identical to those observed in liquid drops and that the presented model can be used to demonstrate the polygonal oscillations in perturbed liquid drops. However, it must be reemphasized that our spring-mass network is a simple model to imitate polygonal oscillations in liquid drops. Therefore, it is not necessary that the shapes of the network shown in \cref{2}((c) and (d)) will be observed in liquid drops. Hence, only a qualitative comparison between the model and the experimental systems can be made. It also needs to be pointed out that for demonstration purposes, the interlayer connections between masses are not shown in \cref{3}; these connections, however, are considered in the simulations.

Liquid drops undergoing polygonal oscillations normally follow the relation $f_n=f_p^{'}/2$. However, in a recent work \cite{Singla}, relation $f_n=f_p^{'}/4$ was reported for the first time in a perturbed liquid drop. In the spring-mass network, this behavior is demonstrated in \cref{3} where time series (a) and the frequency spectra (b) of the masses of the network are shown. Moreover, extending the idea that is established between \cref{2}(b) and \cref{2}(c), it can be stated that in the present case, there will be four prominent inverted polygons that will appear in every cycle of the perturbation as one full oscillation of the network completes. These four prominent polygons (along with the arrows indicating the order of appearance of shapes) are shown in \cref{3}(c) (multimedia view). To reiterate, oscillations in this case are also symmetric (see \cref{sec3}).

\begin{figure}[ht!]
\includegraphics[width=8cm,height=4cm]{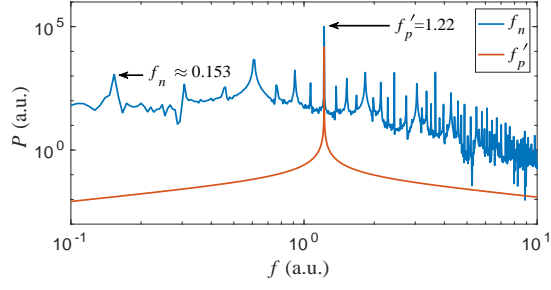}
\caption{Frequency spectra of a mass from the outermost layer of the network (blue curve) and the perturbation (orange curve) for $a'=7.12$, $f_p^{'}=1.22$.}
\label{p8}
\end{figure}

\begin{figure}[ht!]
\includegraphics[width=8cm,height=8cm]{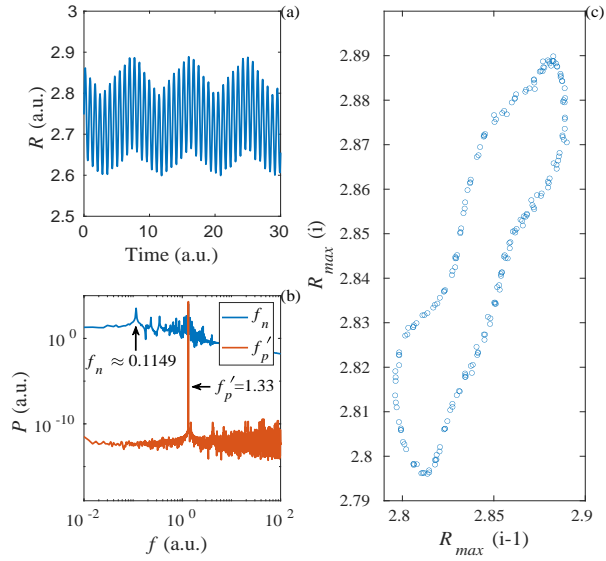}
\caption{(a) Time series of the radii of one mass from the outermost layer of the network ($a'=0.5$, $f_p^{'}=1.33$), (b) frequency spectra of a mass from the outermost layer (blue curve) and the perturbation (orange curve),  and (c) return map of the time series showed in (a).}
\label{quasi}
\end{figure}

Other scenarios of oscillations in the perturbed network with different relations between $f_n$ and $f_p^{'}$ are reported in \cref{p8} and \cref{quasi}. \cref{p8} shows the frequency spectra of the time series of radii of one of the masses from the outermost layer and the perturbation signal; in this case, $f_n=f_p^{'}/8$ is observed. In \cref{quasi}, a scenario where $f_n$ and $f_p^{'}$ are related by irrational numbers is demonstrated; the dynamics of the network in this case is called quasi-periodic. The time series of the radii of a mass of network executing quasi-periodic dynamics is shown in \cref{quasi}(a) and its frequency spectra in \cref{quasi}(b). From \cref{quasi}(b) it can observed that $f_n\approx f_p^{'}/11.6$. Of course, getting an irrational ratio between $f_n$ and $f_p^{'}$ in simulations is nonviable. Therefore, to confirm the quasi-periodic nature of the time series shown in \cref{quasi}(a), its return map is presented. For this purpose, all the maximas of the time series (in \cref{quasi}(a)) are first calculated, and then $i^{th}$ maxima is plotted against $(i-1)^{th}$ maxima on \cref{quasi}(c). The loop-like structure of the return map indicates that the time series shown in \cref{quasi} is quasi-periodic in nature. Polygonal oscillations identical to those shown in \cref{2} and \cref{3} have been reported earlier in experimental systems. However, those of \cref{p8} and \cref{quasi}, or any other relation between $f_n$ and $f_p^{'}$ to the best of our knowledge, have not been reported. Therefore, it will be interesting to see if under appropriate experimental conditions such results can also be observed in perturbed liquid drops.

\begin{figure}[ht!]
\includegraphics[width=8cm,height=8cm]{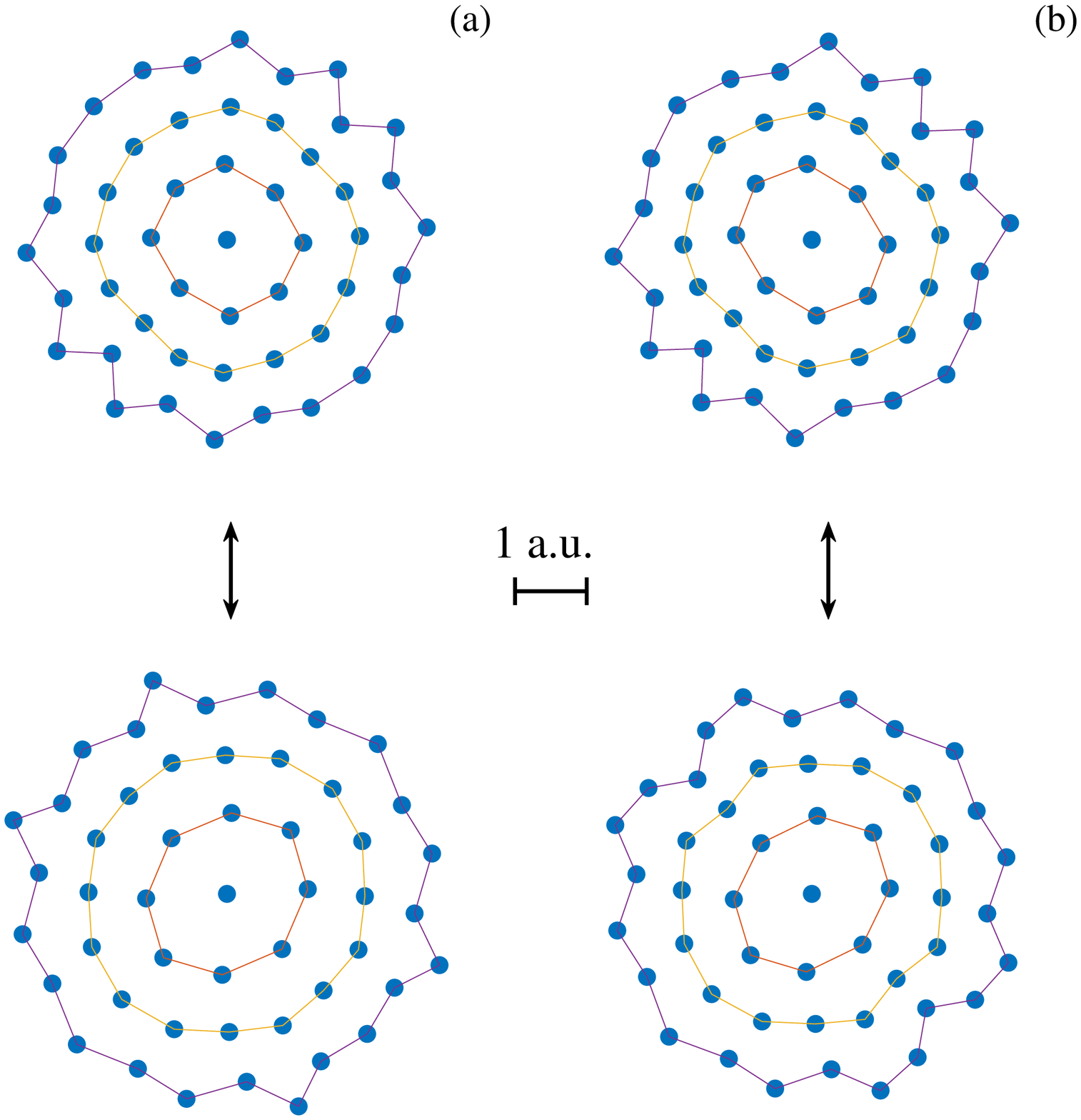}
\caption{Inverted polygonal shapes for asymmetric oscillations in the network for (a) $a'=2$, $f_p^{'}=1.8$ (multimedia view) and (b) $a'=2.92$, $f_p^{'}=1.22$ (multimedia view).}
\label{asym}
\end{figure}

In \cref{asym}(a) (multimedia view) and \cref{asym}(b) (multimedia view), two examples are shown where oscillations are asymmetric. To reiterate, these oscillations are called asymmetric as the masses are not equally distributed among four quadrants of the $x'-y'$ space (see \cref{sec3}). For the results shown on this figure, $f_n=f_p^{'}/2$. In experimental systems, this behavior is observed when the drop oscillates in the shape of an ellipse or any polygon with odd number of lobes \cite{Dinesh}.

\begin{figure}[ht!]
\includegraphics[width=8cm,height=8cm]{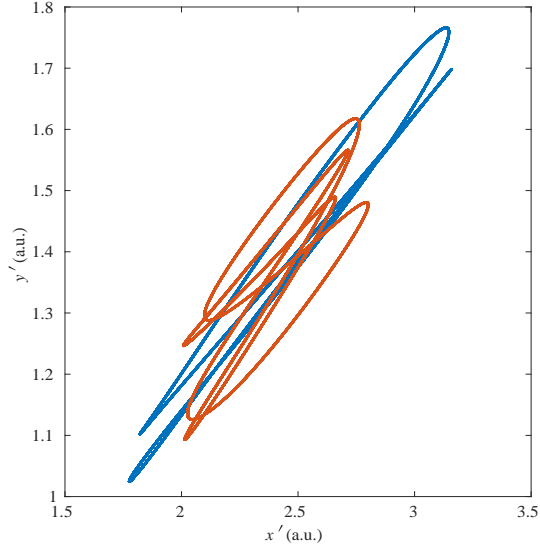}
\caption{$y'$ vs. $x'$ plot of one mass from the outermost layer of the network exhibiting polygonal oscillations. Blue curve: $a'=3.2$, $f_p^{'}=1.22$, orange curve: $a'=2.2$, $f_p^{'}=1.38$.}
\label{rot}
\end{figure}

In \cref{rot}, $y'$ vs. $x'$ curves of a mass from the outermost layer of the network are shown for two different values of the perturbation parameters. Curves on this figure represent polygonal oscillations with $f_n=f_p^{'}/2$ (blue) and $f_n=f_p^{'}/4$ (orange). It can be noticed that, along with radial motion, the mass also exhibits tangential motion. This indicates the emergence of symmetry breaking in the network which results in the polygonal oscillations in the network and also in the liquid drops. Moreover, in \cite{Singla}, it is shown that a vertically vibrated Mercury drop executes rotational motion. Results of \cref{rot} indicate that the tangential component in the motion of masses in the network (particles in the liquid drop) is responsible for the rotational motion of the perturbed drop executing polygonal oscillations.

\begin{figure}[ht!]
\includegraphics[width=8cm,height=8cm]{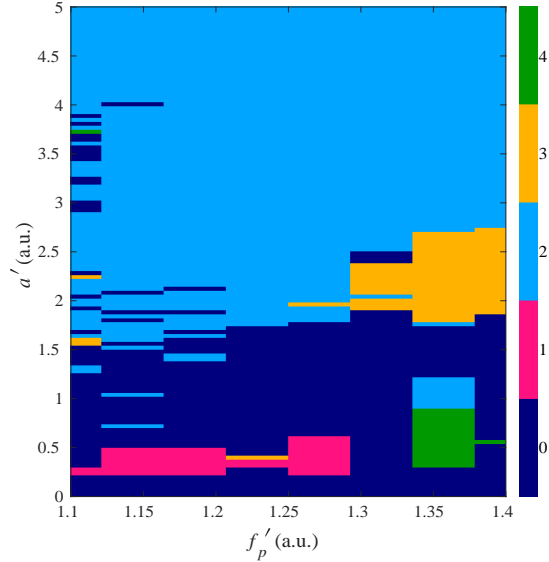}
\caption{Two dimensional parameter space plot representing numerical relations between $f_n$ and $f_p^{'}$ for different values of $a'$ and $f_p^{'}$. 0$\rightarrow$ aperiodic, 1$\rightarrow$ $f_n=f_p^{'}$, 2$\rightarrow$ $f_n=f_p^{'}/2$, 3$\rightarrow$ $f_n=f_p^{'}/4$, and 4$\rightarrow$ quasi-periodic dynamics.}
\label{bif}
\end{figure}

In \cref{bif}, a bifurcation diagram representing different dynamics of the model given by \cref{eq4} is presented. It represents the dynamics in the network on the basis of numerical relation between $f_n$ and $f_p^{'}$. Frequency ($f_p^{'}$) and amplitude ($a'$) of the perturbation are varied to obtain this diagram. On the diagram (and also on the colorbar), the region marked with number 0 corresponds to the situations where numerical relation between $f_n$ and $f_p^{'}$ could not be established. This is due to the fact that at smaller values of $a'$, the amplitude of oscillations in the network remains insignificant. This makes the relation between the frequencies difficult to be established. However, as $a'$ starts to augment, before stable polygonal oscillations are observed in the network, it exhibits switching between unstable polygonal oscillations. Regions marked with (1) represent $f_n=f_p^{'}$, (2) $f_n=f_p^{'}/2$, (3) $f_n=f_p^{'}/4$, and (4) quasi-periodic oscillations. It needs to be mentioned here that the boundaries between two regions of different colors might have small errors. This is due to the fact that in some cases boundaries between two regions are hard to demarcate and the transition from one state to the other is difficult to be identified. The above mentioned bifurcations can also be observed experimentally in perturbed liquid drops {\em i.e.} at small $a'$ or $f_p{'}$, $f_n=f_p^{'}$ is observed and increasing any of these parameters results in polygonal oscillations in the drop. The fact that $f_n$ and $f_p$ are related with both rational and irrational numbers at different perturbation parameters indicates that different frequency locked regions on \cref{bif} are the Arnold tongues of the perturbed network.

\section{Conclusions and Discussions}
A numerical model to simulate polygonal oscillations similar to those observed in a perturbed liquid drop is presented. For this purpose, a simple network consisting of springs and masses was designed and forced with a periodic signal. To facilitate the choice of the network parameters, the model was nondimensionalized which reduced all of its parameters to a single quantity ($b^{'}$). To obtain different results, $a'$ and $f_p^{'}$ were controlled. Phenomena like polygonal oscillations, different numerical relations between $f_n$ and $f_p^{'}$, symmetric and asymmetric oscillations, tangential component in the motion of the masses were observed in the network. Most of these phenomena can also be observed experimentally in a perturbed liquid drop. This makes the proposed model a suitable system for further theoretical investigations in the field of polygonal oscillations in perturbed liquid drops.

Although several similarities are observed between the results of perturbed liquid drops and those of the presented model, there are some results that are exclusive either to the experiments or the proposed model. For example, in \cite{Singla} rotation of a vertically vibrated Mercury drop is reported. In the present work, the masses possessed a tangential component in their motion, however, complete rotation of the network was not observed. In experimental systems, tangential motion of the masses can also be verified by using Particle Induced Velocimetry (PIV). Secondly, in \cref{p8} and \cref{quasi} different numerical relations between $f_n$ and $f_p^{'}$ were provided, however, there is no experimental verification of these relations yet. Masses in the present case were constrained to move on a horizontal plane. This precluded the observation of Faraday waves which, in experimental systems, are present on the surface of liquid drops.

To reemphasize, the presented model in this work does not incorporate actual fluid dynamical mechanisms that take place during polygonal oscillations in a perturbed liquid drop. A computational model based on different hydrodynamical forces in this system will be more complex, require numerical computation of partial differential equations (for e.g. Navier-Stokes equation), and certainly be time consuming. On the contrary, the proposed model is developed by considering the dynamical mechanisms for polygonal oscillations as spring forces. Nonetheless, the model is able to demonstrate most of the features of polygonal oscillations in a perturbed liquid drop. Therefore, it must be treated just as an alternate simple approach to observe polygonal oscillations numerically. This model can also be used to explore experimental systems where other phenomena in liquid drops are studied, for e.g. bouncing of a drop on a rigid surface.

\section{Acknowledgement}
Authors acknowledge financial support from CONACyT project CF140606. 

\section{Data Availability}
All results can be obtained by simulating the model represented by \cref{eq4} at different parameter values.

\appendix
\section{~}\label{app1}
\renewcommand{\theequation}{A\arabic{equation}}
Considering $t=t_0t^{'}$, and ${\bf x_i}=x_0{\bf x_i'}$ in \cref{eq3} and using \cref{eq1} and \cref{eq2} we get:

\begin{align}
m\frac{x_0}{t_0^2}\frac{d^2{\bf x_i'}}{dt^{'2}}&=\sum_{j=1}^N A_{ij}\Big[-kx_0\frac{((x_i'-x_j')^2+(y_i'-y_j')^2)^{1/2}}{\delta}+\frac{k}{x_0^2}\frac{\delta^2}{[(x_i'-x_j')^2+(y_i'-y_j')^2]}\nonumber\\
&-\frac{bx_0^3}{t_0^3}((v_{x_i}'-v_{x_j}')^2+(v_{y_i}'-v_{y_j}')^2)^{3/2}\Big]{\bf c_{\phi_{ij}}}+a\mathrm{sin}(2\pi f_pt_0 t^{'}){\bf c_{\theta_i}},\label{eq5}\\
\frac{d^2{\bf x_i'}}{dt^{'2}}&=\sum_{j=1}^N A_{ij}\Big[-\frac{kt_0^2}{m}\frac{((x_i'-x_j')^2+(y_i'-y_j')^2)^{1/2}}{\delta}+\frac{kt_0^2}{mx_0^3}\frac{\delta^2}{[(x_i'-x_j')^2+(y_i'-y_j')^2]}\nonumber\\
&-\frac{bx_0^2}{mt_0}((v_{x_i}'-v_{x_j}')^2+(v_{y_i}'-v_{y_j}')^2)^{3/2}\Big]{\bf c_{\phi_{ij}}}+\frac{at_0^2}{mx_0}\mathrm{sin}(2\pi f_pt_0 t^{'}){\bf c_{\theta_i}}.\label{eq6}
\end{align}

Here, ${\bf c_{\phi_{ij}}}=\{\mathrm{cos}\phi_{ij},\mathrm{sin}\phi_{ij}\}$ is the unit vector in the direction of the restoring and damping forces of the springs; $\phi_{ij}$ being the angle which the spring joining $i^{th}$ and $j^{th}$ masses make with the $x'$-axis. As mentioned in \cref{sec2}, ${\bf c_{\theta_i}}=\{\mathrm{cos}\theta_i,\mathrm{sin}\theta_i\}$ represents the unit vector in the direction of the perturbation.  Considering $t_0=(m\delta/k)^{1/2}$, and $x_0=\delta$, \cref{eq6} becomes:

\begin{align}
\frac{d^2{\bf x_i'}}{dt^{'2}}&=\sum_{j=1}^N A_{ij}\Big[-((x_i'-x_j')^2+(y_i'-y_j')^2)^{1/2}+\frac{1}{(x_i'-x_j')^2+(y_i'-y_j')^2}\nonumber\\
&-b^{'}((v_{x_i}'-v_{x_j}')^2+(v_{y_i}'-v_{y_j}')^2)^{3/2}\Big]{\bf c_{\phi_{ij}}}+a'\mathrm{sin}(2\pi f_p^{'} t^{'}){\bf c_{\theta_i}},\label{eq7}\\
&=\sum_{j=1}^N A_{ij}\Big[{\bf F^{'}_{r_{ij}}}+{\bf F^{'}_{d_{ij}}}\Big]+a'\mathrm{sin}(2\pi f_pt^{'}){\bf c_{\theta_i}}.\label{eq8}
\end{align}

In \cref{eq7} and \cref{eq8}, $b^{'}=bk^{1/2}(\delta/m)^{3/2}$, $a'=a/k$, $f_p^{'}=f_pt_0$, and  ${\bf F_{r_{ij}}'}$ and ${\bf F_{d_{ij}}'}$ are the nondimensionalized restoring and damping forces.

\begin{figure}[ht!]
\includegraphics[trim=0cm 2cm 0cm 1cm,clip=true,width=10cm,height=7cm]{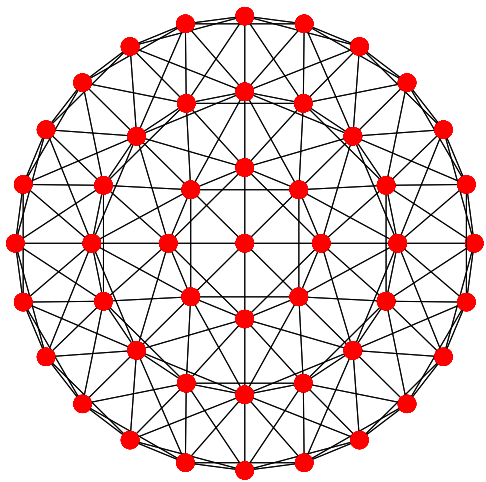}
\caption{Schematic representation of the spring-mass network with increased number of connections for higher stability to observe polygonal oscillations. Red circles represent masses and black lines represent springs.}
\label{network}
\end{figure}

\section{~}\label{app2}
The results of polygonal oscillations have been found to be highly sensitive to the number of masses that are present in each layer of the network. It is observed that if the number of masses increases, the results deviate from what is presented earlier. This problem can be reduced to some extent by introducing more connections between the masses of the network. One of such possibilities is to connect every mass of a layer to their second neighbors from the same layer. A schematic diagram of this network is shown in \cref{network}. This idea can be extended up to a certain limit by connecting every mass to their subsequent neighbors to bring more stability to the network. Increasing interlayer connections can also be explored to this end.

\bibliography{biblio}
\end{document}